\documentclass[showpacs,showkeys,aip,apl,reprint,]{revtex4-1}
\usepackage{graphicx,amsmath,natbib}
\usepackage[latin1]{inputenc}
\begin{document}

\title{Direct determination of the surface termination in full-Heusler alloys by means of low energy electron diffraction}
\author{Jan-Peter W{\"u}stenberg}
\email{jpwuest@physik.uni-kl.de}
\affiliation{Department of Physics and Research Center OPTIMAS, University of Kaiserslautern, Erwin-Schroedinger-Str. 46, 67663 Kaiserslautern, Germany}
\author{Takayuki Ishikawa}
\author{Masafumi Yamamoto}
\affiliation{Division of Electronics for Informatics, Graduate School of Information Science and Technology, Hokkaido University, Sapporo 060-0814, Japan}
\author{Christian Herbort}
\author{Martin Jourdan}
\affiliation{Institut für Physik, Johannes Gutenberg-University, Staudinger Weg 7, 55128 Mainz, Germany }
\author{Martin Aeschlimann$^1$}
\author{Mirko Cinchetti$^1$}
\noaffiliation

\pacs{68.35.Bd, 61.05.jh, 68.35.Dv, 68.55.at}
\keywords{LEED, low energy electron diffraction, Heusler alloy, CMS, CCFA, MgO, tunneling junction, surface termination, sputtering, disorder, spin polarization}

\begin{abstract}
The performance of Heusler based magnetoresistive multilayer devices depends crucially on the spin polarization and thus on the structural details of the involved surfaces. Using low energy electron diffraction (LEED), one can non-destructively distinguish between important surface terminations of Co$_2$XY full-Heusler alloys. We present an analysis of the LEED patterns of the Y-Z ,the vacancy-Z, the Co and the disordered B2 and A2 terminations. As an example, we show that the surface geometries of bulk $L2_1$ ordered Co$_2$MnSi and bulk B2 disordered Co$_2$Cr$_{0.6}$Fe$_{0.4}$Al can be determined by comparing the experimental LEED patterns with the presented reference patterns.
\end{abstract}

\maketitle

Heusler compounds are intermetallic X$_2$YZ alloys crystallizing in the $L2_1$ structure, where X and Y atoms are transition metals and Z is a main group element. 

Many Heusler compounds, in particular those based on $X=Co$ are predicted to be half metallic ferromagnets with high Curie temperatures, promising a performance boost in spintronics devices due to their predicted 100\,\% spin polarization at the Fermi energy \cite{Kubler84}. However, despite good device performance~\cite{Ishikawa09}, a complete surface spin polarization has not been demonstrated yet.

The $L2_1$ structure consists of four interpenetrating fcc sublattices lined up regularly in the sequence X-Y-X-Z along the [111] direction of the cubic unit cell. Another way to describe this structure is a simple cubic lattice of X atoms (with half the lattice constant of the $L2_1$ unit cell), with center positions alternately occupied by Y and Z atoms. 

In bulk crystals, structural disorder has been shown to be a source of low spin polarization, particularly in the case of Co antisite disorder~\cite{Picozzi04,Miura04}. Considering surfaces and interfaces with tunneling barriers, ordering, termination and stoichiometry play a decisive role for the electron spin polarization and thus have to be known precisely. Experiments have shown that spin polarization and TMR ratios depend strongly on the annealing temperature~\cite{Ishikawa09} as well as on the details of the surface structure and morphology~\cite{Wustenberg09,Herbort09,Herbort09a,Cinchetti07}. For Co$_2$MnSi (CMS), theoretical calculations predict four stable surface terminations: Mn-Mn, Mn-Si, Si-Si and vacancy-Si. All stable surface configurations - apart from the Mn-Mn termination - are predicted to reduce the surface spin polarization by the appearance of surface states crossing the Fermi level within the bulk minority band gap~\cite{Hashemifar05}. For CMS/MgO junctions, only Co-Co and Mn-Si interface layers have been predicted to be stable under equilibrium conditions~\cite{Hulsen09,Miura08}. Recent experiments using high-angle annular dark-field scanning transmission electron microscopy suggest that the interface layer is formed by the layer next to Co rather than by a Co-Co layer ~\cite{Miyajima09}, including some degree of atomic disorder within the first two atomic layers. 

In this article, we use low energy electron diffraction (LEED) in order to obtain information about the surface crystal structure. The large scattering cross section of slow electrons ($E_{kin}=20\,eV-500\,eV$) restricts interference to within the first atomic layers, thus almost completely relaxing the 3rd Laue condition perpendicular to the surface. Ideally, only reflexes corresponding to 2D reciprocal lattice vectors $\mathbf{G_{hk}}$ in the surface plane are observed in the diffraction pattern. Additional information can be deduced from the shape of the spot intensity profile, the occurrence of extra spots or lines or from the energy dependence of the diffraction pattern, using kinematic theory.

We have used three different (100) surfaces, namely (1) a surface of a 2-nm-thick MgO tunneling barrier deposited on a 50-nm-thick CMS layer, (2) a surface of a 50-nm-thick CMS layer with bulk L2$_1$ order exposed by the removal of 3nm of MgO and Al oxide, and (3) a surface of a 100-nm-thick Co$_2$Cr$_{0.6}$Fe$_{0.4}$Al (CCFA) layer with bulk B2 disorder exposed by the removal of a 4-nm-thick Al oxide cap.

The layer structure for the MgO barrier sample (1) was as follows: (from substrate side) MgO buffer(10\,nm)/CMS(50\,nm)/MgO barrier(2\,nm), grown on a MgO(100) substrate. The MgO barrier was grown epitaxially at room temperature (RT) by e-beam evaporation. The underlying CMS film exhibited Co$_2$Mn$_{0.69}$Si$_{1.01}$ composition and was annealed to 600\,°C for 15 min before the deposition of the barrier. The sample was immersed in a solvent in order to avoid hygroscopic effects during transport to the LEED vacuum system. The CMS film (2) was deposited by magnetron sputtering on MgO(100)/MgO\,(10\,nm) and annealed to 600\,°C for 15\,min. The film composition was Co$_2$Mn$_{0.91}$Si$_{0.93}$ (Ref.~\onlinecite{Ishikawa08}). It was capped by 2\,nm of MgO and 1\,nm of Al oxide. The CCFA film (3) was deposited at room temperature by rf sputtering on an MgO buffered MgO(100) substrate and annealed to 550\,°C. A 4\,nm Al oxide capping was used for protection~\cite{Herbort09a}.

After introducing into the vacuum system the capping layers of the CMS (2) and CCFA (3) films were removed by 500\,eV $Ar^{+}$ ion sputtering ($6\,\mu A$ sample current, 60\,° angle of incidence) and subsequent annealing to 550\,°C. Auger spectra showed no sign of Al or Mg oxides on the surface after this treatment. The MgO barrier (1) was annealed to 500\,°C until the solvent was desorbed completely from the surface.

LEED patterns of the sample surfaces were obtained by means of an Omicron 3-grid SpectaLEED system after the cleaning procedure described above. From the ratio of spot width (Lorentz FWHM) and reciprocal lattice vector an instrumental transfer width~\cite{Ertl85} of at least 12\,nm was determined at a clean Cu(100) reference surface at $E=80$\,eV.
 

In kinematic LEED theory, the intensity pattern of a perfectly flat surface is given by
\begin{equation}
I(\mathbf{k_{diffr}},\mathbf{k_0})\propto G^2 F_{hk}^2 ,
\end{equation}
where the lattice factor $G$ describes the possible spot positions according to the 2D Laue condition $\mathbf{k_{diffr}}-\mathbf{k_0}=\mathbf{G_{hk}}$, modified by the interference $F_{hk}$ caused by the inner atomic structure of the unit cell \cite{Ertl85}. Our analysis is based on the systematic spot intensity modifications introduced by the different surface terminations via the structure factor $F_{hk}$. 

Given orthogonal surface lattice vectors $\vec{a}_1$,$\vec{a}_2$ spanning the unit cell and $s$ atomic positions $\vec{r}_j$ with respective scattering factors $f_j$, the intensity modifications for a spot $(h,k)$ in the surface diffraction pattern can be written as~\cite{Ertl85}

\begin{equation}
\label{eq_fhk}
	F_{hk}=\sum_{j=1}^{s}{f_j \,e^{2\pi i(h x_j+k y_j)}}.
\end{equation}

\begin{figure}[t]
\includegraphics[width=0.6\linewidth]{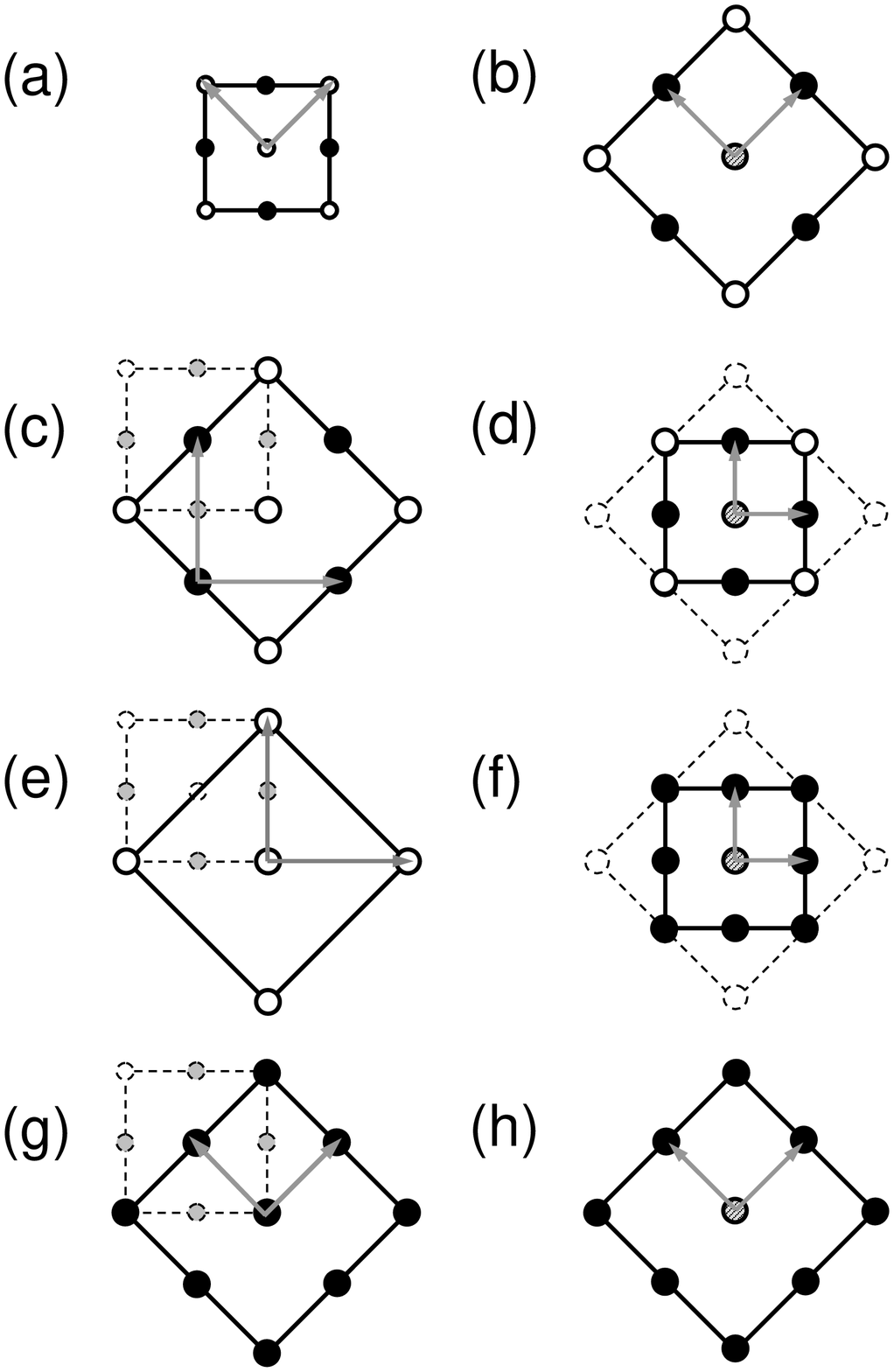}
\caption{Real space (left) and reciprocal lattice (right) for different surface geometries: (a),(b) MgO surface, (c),(d) $L2_1$ Y-Z terminated surface, (e),(f) $L2_1$ Y vacancy, (g),(h) B2 (or X$_2$) termination. In real space different circles represent different atom species, in reciprocal space open (full) circles correspond to high (low) intensities, as determined by the values of $F_{hk}$. Lattice vectors are marked by arrows. The MgO reference is drawn with dashed lines in (c)-(h).}
\label{fig:draw}
\end{figure}

In Fig.~\ref{fig:draw} we have summarized the combinations of the real space surface termination and the expected LEED pattern for the discussed surface configurations. We start with an inspection of the LEED pattern of the MgO(100) barrier (sample (1)) as a reference surface. MgO has a bulk lattice constant $a_{MgO}=4.21\,\AA$ and crystallizes in the NaCl (B1) structure (space group $Fm\bar{3}m$) which can be regarded as a $L2_1$ X$_2$YZ full-Heusler structure without the two X atoms. The MgO(100) surface has the same P4mm symmetry as the Mn-Si terminated CMS(100) surface and is drawn schematically in Fig.~\ref{fig:draw}a. 

\begin{figure}[t]
\centering
\includegraphics[width=\linewidth]{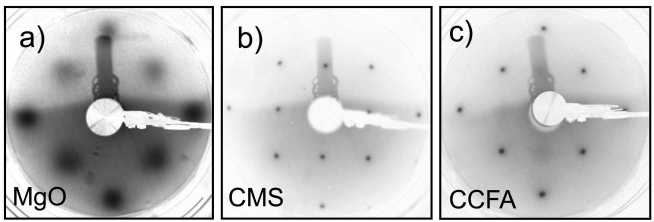}
\caption{LEED patterns of a) the MgO barrier (E=82\,eV), b) CMS (E=57\,eV) and c) CCFA (E=74\,eV).}
\label{fig:pattern}
\end{figure}

The expected LEED pattern is shown in Fig.~\ref{fig:draw}b. The [010] axis of the MgO unit cell is aligned in horizontal direction and coincides with the edges of the quadratic sample in the measurements. 
We observe a cubic diffraction pattern (Fig.~\ref{fig:pattern}a), as reported for cleaved as well as chemically polished MgO(100) surfaces~\cite{Fahsold01,Herbort09}. The apparent rotation of the reciprocal unit cell is caused by the fact that due to the structure factor $F_{hk}$ only reflexes corresponding to reciprocal lattice vectors (h,k) with h \textit{and} k even or odd survive. Equivalently this result can be obtained by noting that on the surface we can chose a smaller surface unit cell (marked by arrows in Fig.~\ref{fig:draw}a) which is a square with the atomic positions $((0,0),(1/2,1/2))$. Thus we can apply a 2D version of the fcc extinction rules known from x-ray diffraction:

\begin{gather}
\label{eq:rules}
h+k \,even \Rightarrow F(h,k)=f_{A}+f_{B}\\
\notag h+k \,odd \Rightarrow F(h,k)=f_{A}-f_{B},
\end{gather}
where A and B correspond to the atoms in the unit cell (in this case Mg and O). The arrows in Fig.~\ref{fig:draw} indicate the real space and reciprocal lattice vectors used to construct the patterns according to Eq. (\ref{eq:rules}).

Most Co-based Heusler alloys favor a 45° rotated growth on the MgO(100) surface. The resulting lattice mismatch is relatively small ($-5.1\,\%$ in the case of CMS and $-3.7\,\% $ in the case of CCFA)~\cite{Yamamoto08}, leading to epitaxial smooth surfaces after annealing~\cite{Herbort09a,Kijima06}. In the case of Y-Z termination the unit cell geometry is the same as in MgO (P4mm), but larger by a factor of $\sqrt{2}$ and rotated by 45°, as shown in Fig.~\ref{fig:draw}c.

CMS is known to assume the L2$_1$ structure in the bulk. The lattice constant of CMS is $a_{CMS}=5.65\,\AA\approx \sqrt{2}\,a_{MgO}$. According to calculations~\cite{Hashemifar05} the Mn-Si and vacancy-Si surface terminations of CMS are thermodynamically most likely to occur. The expected LEED patterns for both surface configurations is plotted in Fig.~\ref{fig:draw}d and Fig.~\ref{fig:draw}f. We find a good qualitative agreement with the geometry of the observed LEED pattern of sample (2) (Fig.~\ref{fig:pattern}b). However, according to Fig.~\ref{fig:draw}d the intensity of edge spots (full circles) should be smaller than that of corner spots (open circles) for a Mn-Si termination. The LEED pattern of the investigated CMS sample does not show such a variation of spot intensities. This indicates that one of the atomic form factors might be very small or either Mn or Si atoms have vanished completely from the surface lattice, leaving either Y or Z vacancies behind. This could be caused by an element selective depletion of the surface during the preparation steps, e.g. Ar$^+$ etching. 
Indeed the Ar$^+$ sputtering rates of monoatomic Mn is $2.5$ times the one from Si, favouring a Si rich surface~\cite{NPL05}. We conclude that even though a definite distinction between the Mn-Si terminated surface and the vacancy-Si terminated surface cannot be made at this level of analysis, other configurations, like a Co, B2, Mn-Mn or Si-Si surface termination can be excluded from the observed pattern.  

For B2 disordered, Co terminated or A2 surfaces the rules (\ref{eq:rules}) lead to an extinction of every second spot~\cite{Henzler77}. 
This can be seen for the case of CCFA, which is known to crystallize in the B2 structure. The LEED pattern is shown in the right panel of Fig.~\ref{fig:pattern}c. Due to the random intermixing of Y and Z atoms the surface is effectively monoatomic. As expected from the illustration in Fig.~\ref{fig:draw}h the pattern is expanded and rotated by 45° with respect to the pattern of the Y-Z or vacancy-Z terminated surfaces.

In conclusion, we have shown that by using a standard surface science tool it is possible to distinguish between technologically important surface terminations of full Heusler alloys with different bulk structure. For $L2_1$ order, Y-Z and vacancy-Z terminations can be distinguished from patterns caused by bulk B2, A2 or $L2_1$ Co terminated surfaces. For example, depending on the background information available, LEED can be used in an $L2_1$ structure to separate easily a Y-Z from a Co terminated surface. By closer inspection of the spot intensities one can draw conclusions about details on the surface termination which would otherwise be hard to obtain. Therefore LEED analysis should be included in a multitechnique approach for investigating the electronic properties of Heusler surfaces.

Financial support of the DFG FOR-559 "New materials with high spin polarization" is gratefully acknowledged.


\begin{thebibliography}{10}
\providecommand{\url}[1]{\texttt{#1}}
\providecommand{\urlprefix}{URL }

\bibitem{Kubler84}
J.~K\"ubler, Physica B+C \textbf{127}, 257 (1984).

\bibitem{Ishikawa09}
T.~Ishikawa, N.~Itabashi, T.~Taira, K.~Matsuda, T.~Uemura, and M.~Yamamoto, J.
  Appl. Phys. \textbf{105}, 07B110 (2009).

\bibitem{Picozzi04}
S.~Picozzi, A.~Continenza, and A.~J. Freeman, Phys. Rev. B \textbf{69}, 094423
  (2004).

\bibitem{Miura04}
Y.~Miura, K.~Nagao, and M.~Shirai, Phys. Rev. B \textbf{69}, 144413 (2004).

\bibitem{Wustenberg09}
J.-P. W\"ustenberg, J.~Fischer, C.~Herbort, M.~Jourdan, M.~Aeschlimann, and
  M.~Cinchetti, J. Phys. D: Appl. Phys. \textbf{42}, 084016 (2009).

\bibitem{Herbort09}
C.~Herbort, E.~{Arbelo Jorge}, and M.~Jourdan, Appl. Phys. Lett. \textbf{94},
  142504 (2009).

\bibitem{Herbort09a}
C.~Herbort, E.~Arbelo, and M.~Jourdan, J. Phys. D: Appl. Phys. \textbf{42},
  084006 (2009).

\bibitem{Cinchetti07}
M.~Cinchetti, J.-P. W\"ustenberg, M.~S. Albaneda, F.~Steeb, A.~Conca,
  M.~Jourdan, and M.~Aeschlimann, J. Phys. D: Appl. Phys. \textbf{40}, 1544
  (2007).

\bibitem{Hashemifar05}
S.~J. Hashemifar, P.~Kratzer, and M.~Scheffler, Phys. Rev. Lett. \textbf{94},
  096402 (2005).

\bibitem{Hulsen09}
B.~H\"ulsen, M.~Scheffler, and P.~Kratzer, Phys. Rev. Lett. \textbf{103},
  046802 (2009).

\bibitem{Miura08}
Y.~Miura, H.~Uchida, Y.~Oba, K.~Abe, and M.~Shirai, Phys. Rev. B \textbf{78},
  064416 (2008).

\bibitem{Miyajima09}
T.~Miyajima, M.~Oogane, Y.~Kotaka, T.~Yamazaki, M.~Tsukada, Y.~Kataoka,
  H.~Naganuma, and Y.~Ando, Appl. Phys. Express \textbf{2} (2009).

\bibitem{Ishikawa08}
T.~Ishikawa, S.~Hakamata, K.~Matsuda, T.~Uemura, and M.~Yamamoto, J. Appl.
  Phys. \textbf{103}, 07A919 (2008).

\bibitem{Ertl85}
G.~Ertl and J.~K\"uppers, Low energy electrons and surface chemistry, VCH
  Wiley, Weinsheim, 1985.

\bibitem{Fahsold01}
G.~Fahsold, A.~Priebe, and A.~Pucci, Appl. Phys. A \textbf{73}, 39 (2001).

\bibitem{Yamamoto08}
M.~Yamamoto, T.~Marukame, T.~Ishikawa, K.~I. Matsuda, and T.~Uemura, Adv. in
  Solid State Phys. \textbf{47}, 105 (2008).

\bibitem{Kijima06}
H.~Kijima, T.~Ishikawa, T.~Marukame, H.~Koyama, K.~Matsuda, T.~Uemura, and
  M.~Yamamoto, IEEE Trans. Magn. \textbf{42}, 2688 (2006).

\bibitem{NPL05}
NPL, Sputter yield values, Website (2005),
  \url{http://www.npl.co.uk/nanoscience/surface-nanoanalysis/products-and-serv%
ices/sputter-yield-values}.

\bibitem{Henzler77}
M.~Henzler, Electron Spectroscopy for Surface Analysis, chapter~4, Springer
  Verlag Berlin Heidelberg New York Tokyo, 1977.

\end{thebibliography}
\end{document}